\documentclass[a4paper]{jpconf}
\usepackage{graphicx}
\usepackage{amsmath,amssymb}
\usepackage[labelfont=bf]{caption}
\usepackage{graphicx}
\usepackage{booktabs}
\usepackage{cleveref}

\begin{document}
\title{Feedforward control for wave disturbance rejection on floating offshore wind turbines}

\author{M. Al$^{\boldsymbol{\mathsf{1,3}}}$, A. Fontanella$^{\boldsymbol{\mathsf{2}}}$, D. van der Hoek$^{\boldsymbol{\mathsf{1}}}$, Y. Liu$^{\boldsymbol{\mathsf{1}}}$, M. Belloli$^{\boldsymbol{\mathsf{2}}}$ and J. W. van Wingerden.$^{\boldsymbol{\mathsf{1}}}$}

\address{$^{\mathsf{1}}$ Delft Center for Systems and Control, Delft University of Technology, Delft, 2628 CD, The Netherlands
$^{\mathsf{2}}$ Mechanical Engineering Department, Politecnico di Milano, Milano, Via~La~Masa~1, 20156, Italy 
$^{\mathsf{3}}$ sowento GmbH, Donizettistr. 1A, 70195 Stuttgart, Germany}

\ead{al@sowento.com}

\begin{abstract}
Floating offshore wind turbines allow wind energy to be harvested in deep waters. However, additional dynamics and structural loads may result when the floating platform is being excited by wind and waves.
In this work, the conventional wind turbine controller is complemented with a novel linear feedforward controller based on wave measurements. 
The objective of the feedforward controller is to attenuate rotor speed variations caused by wave forcing. To design this controller, a linear model is developed that describes the system response to incident waves.
The performance of the feedback-feedforward controller is assessed by a high-fidelity numerical tool using the DTU 10MW turbine and the INNWIND.EU TripleSpar platform as references. 
Simulations in the presence of irregular waves and turbulent wind show that the feedforward controller effectively compensates the wave-induced rotor oscillations. 
The novel controller is able to reduce the rotor speed variance by $26\%$. 
As a result, the remaining rotor speed variance is only $4\%$ higher compared to operation in still water.

\end{abstract}

\section{Introduction} 
A large portion of the wind energy resource is observed offshore and in waters deeper than 50 meters, where it is not feasible to deploy wind turbines by means of traditional bottom-fixed solutions. Floating offshore wind turbines (FOWTs) are a solution to this problem. However, the non-fixed platform results in additional engineering challenges: In particular, the low-frequency modes associated with the platform rigid-body motion and the additional wave forcing may lead to increased fatigue loads and power oscillations. 
The majority of FOWTs use a variable-speed variable-pitch (VS-VP) controller based on generator speed feedback for rotor speed and power regulation. For FOWTs, the control objective is to track the nominal power curve in the presence of the additional forcing due to wind and waves disturbances.
The presence of rigid-body motion modes associated with the floating platform poses additional constraints to the effectiveness of such a control strategy. As has been widely shown \cite{jonkman2008influence}, the collective pitch controller (CPC) may interact with the platform modes resulting in large motions of the floating structure. This is named the negative damping problem (NDP). In order to prevent it, the CPC bandwidth is decreased, trading the capability of rejecting wind and wave disturbances for lower platform motions. 

More advanced control strategies have been considered in previous works in order to improve the power tracking capabilities while being subjected to wind and wave disturbances.
A model-based linear quadratic regulator (LQR) controller was proposed in \cite{Lemmer_2016}. 
LQR has been proved to be an effective way to reject wind disturbances. However, it is still ineffective in rejecting wave-induced effects. 
A more advanced control strategy is non-linear model predictive control (NMPC)  \cite{schlipf2013nonlinear}. Using a simplified model of the FOWT and a full preview of the incoming wind and wave disturbances, an optimisation algorithm calculates the optimal control action. NMPC obtains superior performance in terms of power production variation reduction and load reduction. However, the optimisation problem is computationally too demanding to be solved in real-time. While NMPC can be seen as an upper limit for FOWT controller performance, a more simple control logic is required to calculate the control action in real-time control purposes.

To improve the wind disturbance rejection capabilities, LIDAR assisted feedforward (FF) control is found to be an effective technology \cite{schlipf2015collective}. A FF action, fed by a measurement of the incoming wind field, is added to the control action of a conventional feedback (FB) controller. This FF control logic attenuates the wind excitation more effectively while preserving the stability and simplicity of the FB controller.

The main objective of this work is to develop a similar FF control strategy based on waves, such that wave disturbances can be compensated for in a simple manner and without compromising on realistic implementation.
Particularly, care is given to obtaining an accurate linearized model of the FOWT dynamics, based on a real-time preview of the surface elevation. 
A novel FF controller is formulated based on this linear model and it is shown that adding such a control logic to the standard FB controller improves the performance of the FOWT. 
To verify the compensation capabilities, this paper focuses on compensating wave-induced rotor speed variations. However, the same procedure can be used to compensate other wave-induced disturbances such as platform pitch motion or tower-base loads.
It is already proven that an accuracy surface elevation preview can be obtained \cite{blondel2012reconstruction} by using regular ship radar systems, which are available at relatively low cost. Therefore, in this work, the emphasis is placed on demonstrating the potential of wave-FF control rather than exploiting an implementation of the wave measurement technology.

The remainder of this paper is organised as follows. Section 2 introduces the FOWT that is considered for developing the control strategy. Section 3 presents the linear model used to describe the wave-induced dynamics. Section 4 introduces the control law of the novel controller. In Section 5, the reference FOWT is subjected to high-fidelity simulations and the results are discussed. The paper is concluded in Section 6. 

\section{Case study}
The effectiveness of the wave FF controller is demonstrated via a case study. The INNWIND.EU TripleSpar platform concept is used \cite{INNWIND} together with the DTU 10MW wind turbine \cite{10MWDTU}. The DTU VS-VP controller is used as the baseline (BL) FB controller.

\begin{figure}
\centering
        \includegraphics[width=50mm]{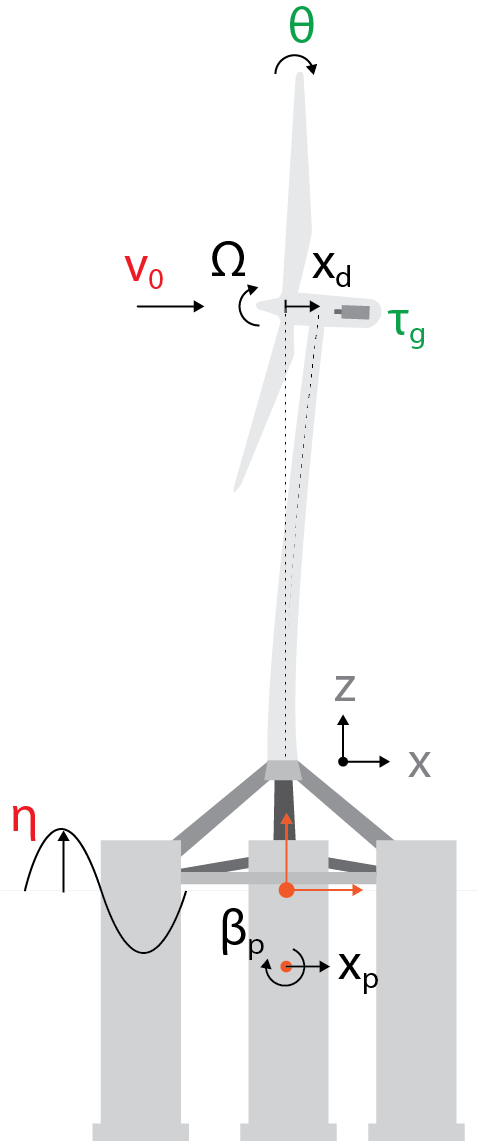}
        \captionof{figure}{Visualisation of the model used in this case study, containing the degrees of freedom (black), the control inputs (green) and the disturbance inputs (red).}
  \label{fig:DOF}
\end{figure}

The dynamic model is reduced to only the most fundamental Degrees of Freedom (DOF), as shown in \cref{fig:DOF}. We consider the platform pitch $\beta_p$, platform surge $x_p$, elastic tower deflection $x_d$ and rotor speed $\Omega$. The controllable inputs are the collective pitch angle $\theta_c$ and the generator torque $\tau_g$. The disturbance inputs are the rotor effective wind speed $v_0$ and the surface elevations $\eta$ at the platform origin. Linear wave kinematics and wave excitation forces were considered for zero-deg wave heading and zero wind-wave misalignment. Wave forces are modelled using potential flow theory.

\section{Linear modelling of the wave disturbance  effects} 
\begin{figure}[h]
    \centering
    \includegraphics[width=0.80\linewidth]{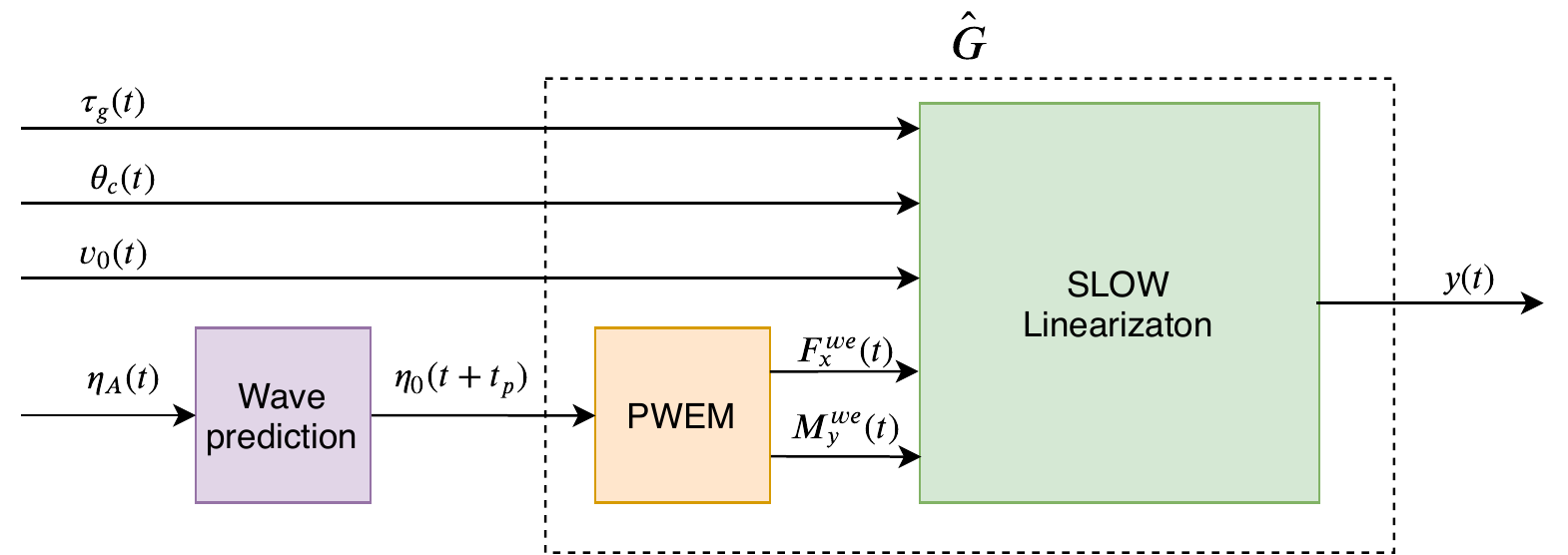}
    \caption{Block diagram of the linear model and its sub-systems.}
    \label{fig:linear-model}
\end{figure}

A linear model of the FOWT allows predicting the system dynamics as a function of disturbance/control inputs, apply linear control engineering theory and develop a linear model-based FF controller. \Cref{fig:linear-model} shows a block diagram representation of the linear approximation model $\hat{G}$ used in this work, consisting of three sub-systems: 
\begin{itemize}
    \item The linearized FOWT dynamics obtained from Simplified Low-Order Wind turbine (SLOW) by \cite{lemmer2020multibody}. In this work, SLOW computes the plant TFs between the outputs $y$ and the inputs: generator torque $\tau_g$, collective blade pitch angle $\theta_c$, wind speed $v_0$ and wave excitation forces $[F_x^{we},\ M_y^{we}]$.
    \item A parametric wave excitation model (PWEM), mapping the surface elevations $\eta$ to wave forces $[F_x^{we},\ M_y^{we}]$.
    \item A surface elevation prediction model that uses the wave elevation measured at time $t$ in point A in front of the FOWT $\eta_A(t)$ to predict the wave elevation at the FOWTs centre of buoyancy at time $t+t_d$, named $\eta_0(t+t_d)$.
\end{itemize}
The linear model is mainly based on SLOW. For more information on SLOW, the reader is referred to \cite{lemmer2020multibody}. The following subsections derive the PWEM and the wave prediction model.

\subsection{Parametric wave-excitation model}
    The PWEM describes the wave forces by wave elevation measurements. Wave forces are introduced in the majority of time-domain simulation potential flow models by means of non-parametric frequency-dependent coefficients, obtained from panel code (e.g. WAMIT) pre-calculations.
    
\begin{figure}[h]
        \centering
        \includegraphics[width=0.8\linewidth]{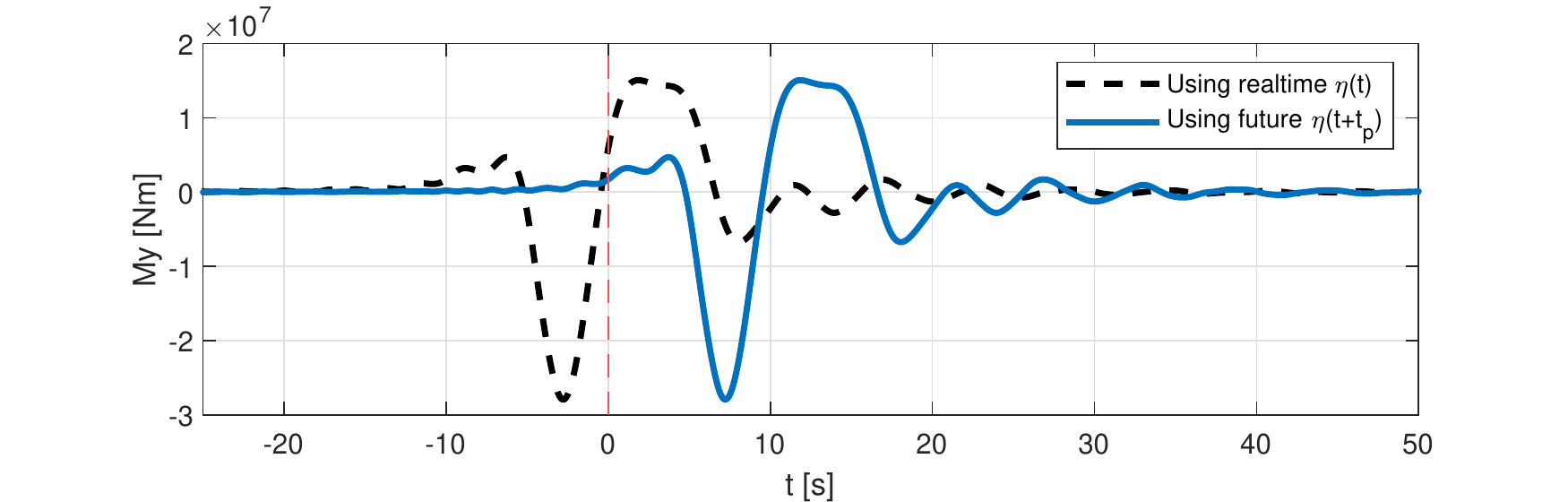}
        \caption{Impulse response of force coefficients, indicating non-causality for the non-delayed force coefficients}
        \label{fig:causality-problem}
    \end{figure}

    The force coefficients represent a non-causal model \cite{Falnes}. This is exemplified in \cref{fig:causality-problem}. The impulse response of the force coefficients, shown by the black dashed line, results in wave forces at negative times. This occurs because wave force coefficient panel codes consider the wave forces to be caused by the wave elevation at the platform centre of buoyancy. In reality, forces are present as soon as the wave impacts the front face of the structure.
    In order to make the model causal, its output response has to be delayed by $t_p$ seconds, where $t_p$ is the smallest possible time for which a wave elevation impulse at time $t=0$ does not result in significant forces at negative times. The response of the causalized system is shown in \cref{fig:causality-problem} in blue, with a time delay of $t_p=10$ seconds.
    
\begin{figure}[h]
        \centering
        \includegraphics[width=\linewidth]{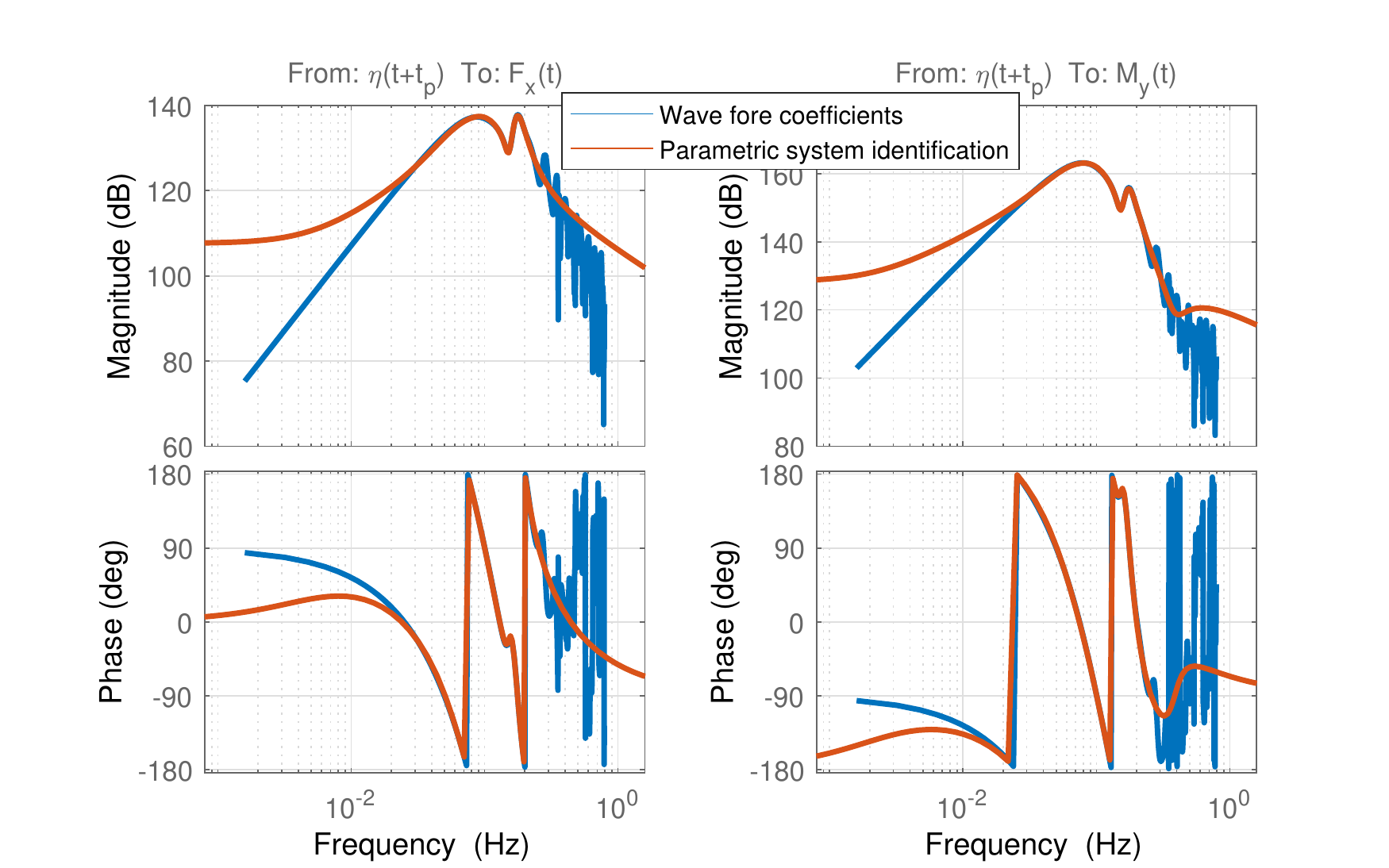}
        \caption{Subspace identification of the non-parametric wave force coefficients. The result denotes the PWEM.}
        \label{fig:system-identification}
    \end{figure}

    Now, a linear time-invariant (LTI) state-space model relating wave elevation to wave forces is obtained from the causalized non-parametric model by means of frequency-domain subspace identification. This parameterization was first proposed in \cite{PWEM} using time-domain system identification. The surge force and pitch moment coefficients of the TripleSpar are visualised in \cref{fig:system-identification} in blue.
    The identification was carried out by means of the N4SID method implemented in Matlab and resulted in a 9th-order single-input multi-output model, shown in \cref{fig:system-identification}. The fit to estimation is found to be $88\%$ and $96\%$ for the surge force and pitch moment respectively, using Akaike's final prediction error (FPE) method. Moreover, the fit is especially good in the wave typical frequency range, from $1/20$ Hz to $1/3$ Hz.

\subsection{Wave prediction model} \label{sec:wave-prediction} 
The PWEM obtained in the previous section is input-output delayed. Thus, if the wave elevation at the platform location at time $t$ is considered for input, the output wave forces are obtained at time $t-t_p$. 
To know forces at the current time $t$, the PWEM must be fed with the wave elevation at the future time $t+t_p$. This information can be extracted from the measurement of surface elevation upstream of the FOWT. The transfer function (TF) that relates the upstream wave elevation measurement $\eta_A(t)$ and the wave elevation at platform location at the future time $\eta_0(t+t_p)$ is obtained from the combination of two TFs.
The first TF puts into relation the surface elevation at two points at the same time instant in deep water:
\begin{equation}
    H(\omega) = \frac{\eta_0}{\eta_A}\exp{\biggl(-i\frac{\omega^2 L}{g}\biggr)},
    \label{eq:place-shift}
\end{equation} 
where $\eta_A$ and $\eta_0$ are the wave elevation at the upstream point and at platform location and $L$ is the distance between the point A and the platform location. The second TF is the expression for a negative time delay $t_p$ in the frequency domain.
\begin{equation}
    P(\omega)=\exp{(i\omega t_p)},\;
    \label{eq:time-delay}
\end{equation}
Thus, combining \cref{eq:place-shift} and \cref{eq:time-delay}, the TF between the upstream and platform wave elevations is obtained:
\begin{equation}
    H_p(\omega)=H(\omega)P(\omega) = \exp{\biggl(i \omega \biggl(t_d - \frac{\omega L}{g}\biggr)\biggr)},
\end{equation}
Causality is obtained if the argument of the exponent is negative. By substituting the wave period $T$ as $\omega=2\pi/T$, requiring causality for waves with a period up to $\overline{T}$ seconds, the minimal measurement distance $L$ in front of the FOWT denotes:
\begin{equation}
    L \geq \frac{g \overline{T} t_p}{2\pi}
\end{equation} 
For this case study, to predict waves up to a period of $\overline{T}=20$ seconds, for $t_p=10$ seconds in advance requires a minimum measuring distance of $L=313$ meters.

\section{Feedforward controller design} 
\begin{figure}[h]
        \centering
        \includegraphics[width=0.63\linewidth]{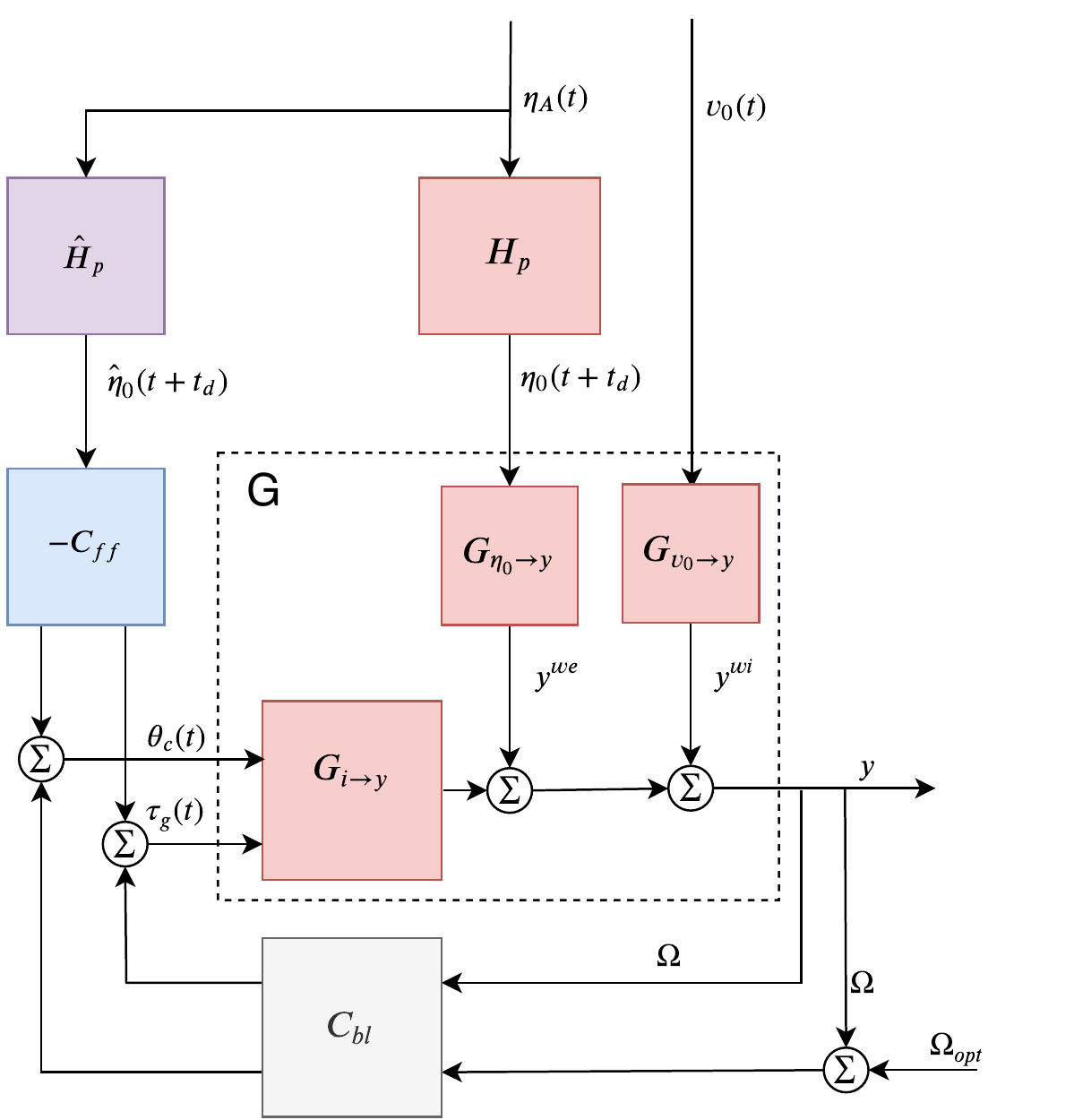}
        \caption{Control logic of the BL controller complemented with the wave-FF controller.}
        \label{fig:control-logic}
    \end{figure}

    \Cref{fig:control-logic} presents the novel control logic in a block diagram. The FOWT plant, named $G$, is controlled by the collective blade pitch angle $\theta_c$ and the generator torque $\tau_g$, based on a measurement of the rotor speed $\Omega$. This control loop is the so-called feedback control loop. Meanwhile, the rotor effective wind speed $v_0$ and surface elevations $\eta_0$ are acting on the same plant $G$, called the disturbances. The description so far is a regular feedback-controlled FOWT block scheme subjected to wind and wave disturbances.
    
    A novel FF controller $C_{ff}$ and a wave predictor $\hat{H}_p$ are added to the regular BL feedback controller.
    The FF controller $C_{ff}$ computes a control action additionally to the feedback controller, by using an upstream measurement of the surface elevation $\eta_A(t)$. The additional control signal will be designed to attenuate the dynamics that are caused by wave disturbances $y^{we}$. A proportional gain $k_{ff}$ is included in the controller to achieve a trade-off between the control objective and the control action.
    
\begin{figure}[h]
        \centering
        \includegraphics[width=0.7\linewidth]{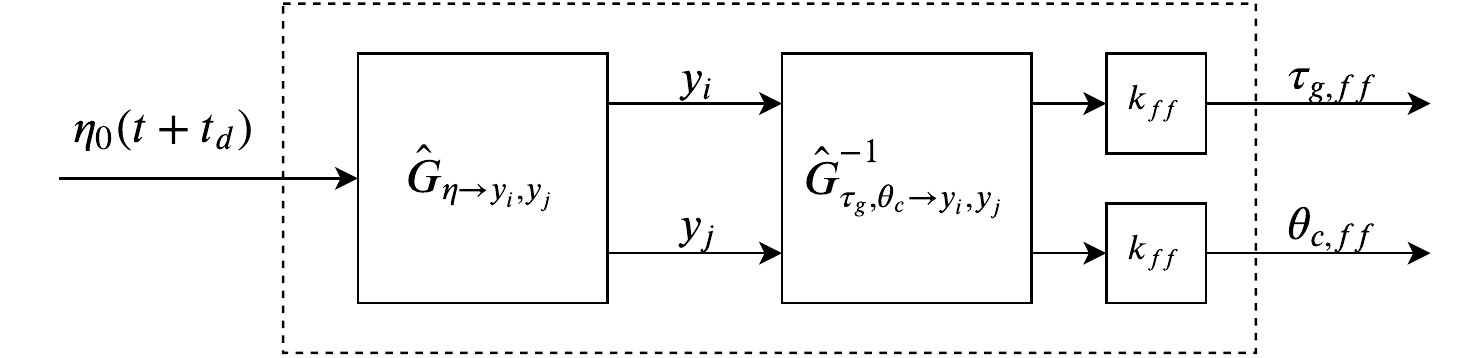}
        \caption{Block diagram of a general FF controller $C_{ff}$ with the objective to compensate the wave-induced effect of two arbitrary system outputs $y_i$ and $y_j$. $k_{ff}$ denotes the proportional gain, allowing the controller to be tuned less aggressive.}
        \label{fig:control-law}
    \end{figure}

    The FF controller $C_{ff}$ is designed using the linear approximation of the effect of the wave disturbance $\hat{G}_\eta$, coupled with an inverse $\hat{G}^{-1}_i$. It computes the additional control inputs for the generator torque $\tau_{g,ff}$ and collective pitch angle $\theta_{c,ff}$, using the control law shown in \cref{eq:FF-pitch}. \Cref{fig:control-law} illustrates the general block diagram of this control law. The design compensates the wave-induced effect of two arbitrary system outputs $y_i$ and $y_j$.
    \begin{equation}
        u_{ff}(s)= -\underbrace{k_{ff} \cdot \hat{G}_{\eta \to \Omega}(s) \cdot \hat{G}_{u_i \to \Omega  }^{-1}(s)}_{C_{ff}(s)} \cdot \eta_{0,p}(s)
        \label{eq:FF-pitch}
    \end{equation}
    
    If the outputs are controllable, the controller can compensate up to two arbitrary system outputs. The performance of the controller depends on the quality of the linear model. To prove the methodology, a FF controller for attenuating rotor-speed variations is designed. When the wind turbine is operating in partial load conditions, the FF controller will act on the generator torque and when in full load conditions on the collective blade pitch angle.
    
\begin{figure}[h]
        \centering
        \includegraphics[width=\linewidth]{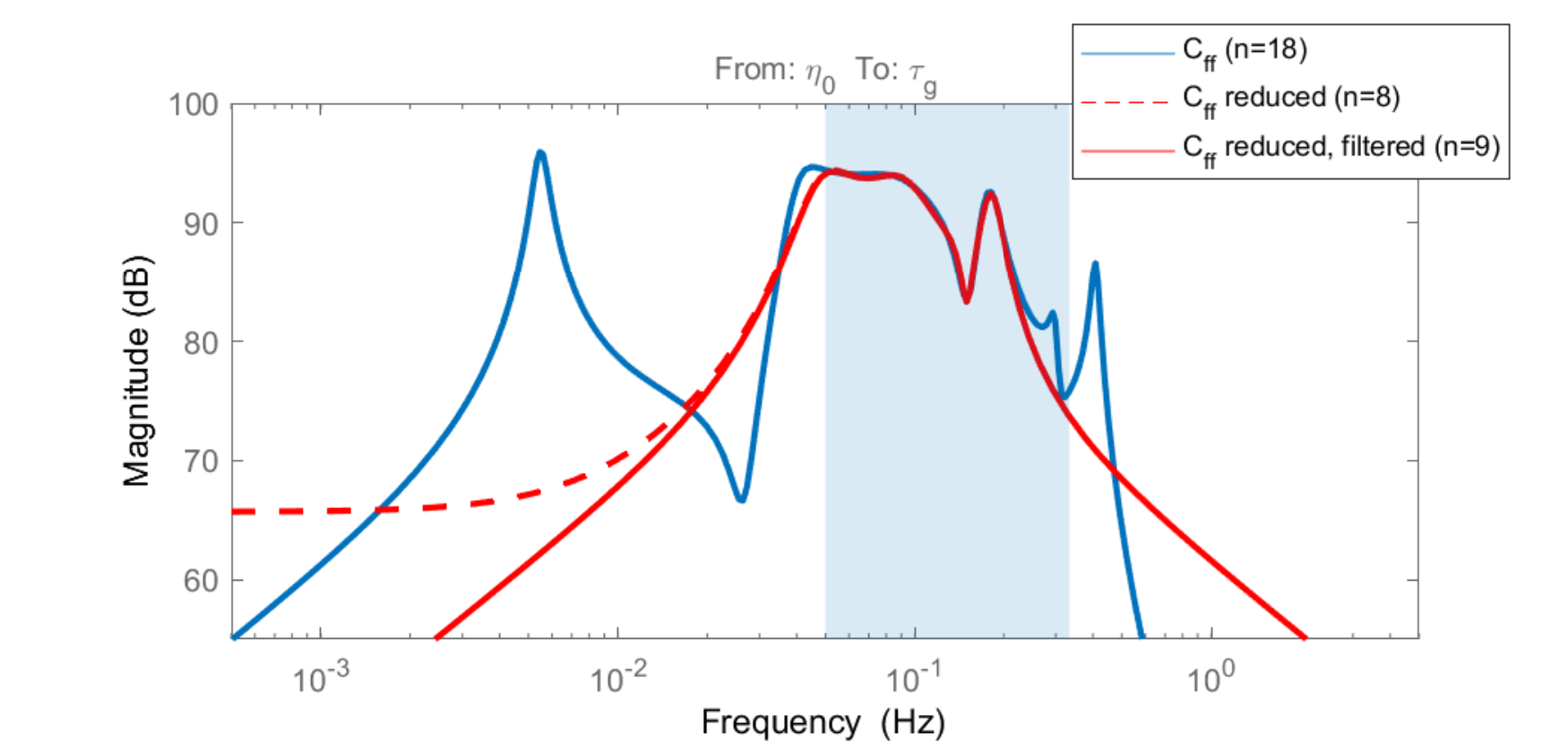}
        \caption{Bode plot of the torque controller and the effect of loop shaping, for operating point $\overline{v}_0=8$ m/s.}
        \label{fig:control-TF}
    \end{figure}

    For this configuration, the outcome of \cref{eq:FF-pitch} is an 18th-order LTI TF. An example of the Bode magnitude plot is shown in \cref{fig:control-TF} for operating point $\overline{v}_0=8$ m/s. 
    Incident ocean waves typically only contain a significant amount of energy for a period of $3>T\geq20$ s. This frequency-range is highlighted in blue. The controller should only compensate for wave responses in this frequency range.
    Because the 18th-order LTI TF also is sensitive to waves of lower frequencies, the order of the original controller is appropriately reduced to an 8th order system (red dashed line) with similar properties in the frequency range of interest. Moreover, a high-pass filter is applied to reduce the sensitivity even further. The final controller is a 9th-order LTI TF, shown by the red solid line.
    
\section{High-fidelity simulations results} \label{sec:results}
To evaluate the performance of the controller, high-fidelity simulations are carried out in FAST v8.16 using the environmental conditions indicated in \cref{tab:load-cases}. The performance of the BL controller is compared to the performance of the same controller extended with the wave-FF controller. Moreover, the BL controller is also simulated without waves to provide an upper-performance limit for wave-FF control.
The control objective is to reduce rotor speed variation. The performance of the two controllers is compared in terms of rotor speed variance ($\Omega$), mean power production ($P$), mean blade pitch action ($\dot{\theta}_c$), tower-base fatigue damage ($M_{ty}$), blade fatigue damage ($M_{b}$) and low-speed shaft fatigue damage ($M_{lss}$). The fatigue damage is measured by a 1-Hz Damage Equivalent Load (DEL).
\begin{table}[h]
    \caption{Selection of load cases based on their occurrence probability, selected from LIFES50+ (DLC1.2).}
        \label{tab:load-cases}
        \centering
        \begin{tabular}{lllll}
        \hline
                    & $v_0 $    & Hs      & Ts      & p       \\
                    & {[}m/s{]} & {[}m{]} & {[}s{]} & {[}-{]} \\ \hline
        Load case 1 & 5         & 1,4     & 7       & 14\%  \\
        Load case 2 & 7,1       & 1,7     & 8       & 24\%  \\
        Load case 3 & 10,3      & 2,2     & 8       & 26\%  \\
        Load case 4 & 13,9      & 3       & 9,5     & 20\%  \\
        Load case 5 & 17,9      & 4,3     & 10      & 11\%  \\
        Load case 6 & 22,1      & 6,2     & 12,5    & 3.8\%  \\
        Load case 7 & 25        & 8,3     & 12      & 0.74\%  \\ \hline
                    &           &         &         & $\approx$100\%  
        \end{tabular}
\end{table}

\begin{figure}[h]
    \centering
    \includegraphics[width=0.9\linewidth]{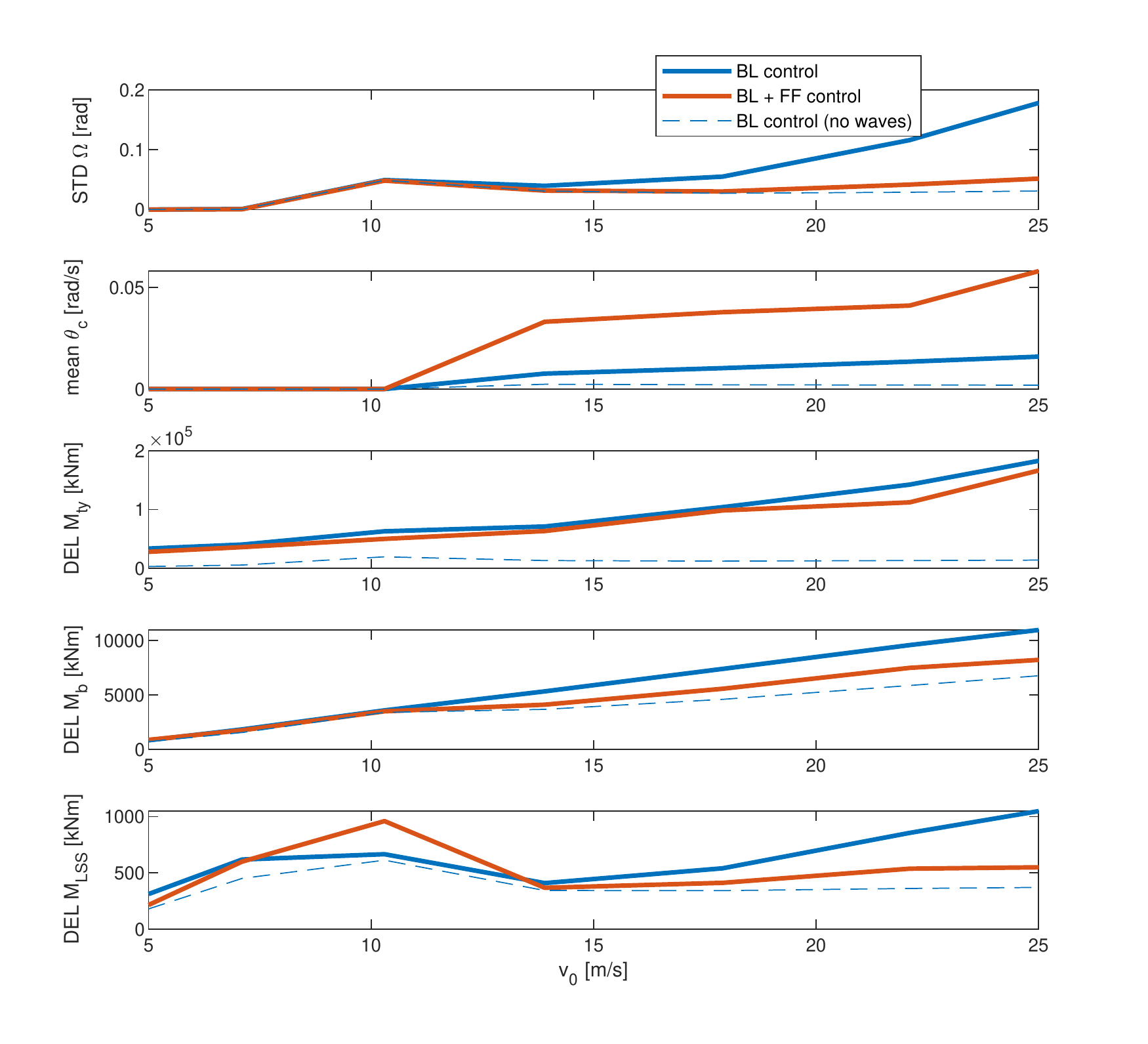}
    \caption{Performance resulting from the high fidelity simulations for each controller and for each load case.}
    \label{fig:results}
\end{figure}

The results demonstrate that wave-FF is an effective control strategy to reject wave-induced rotor speed variations for FOWTs. \Cref{fig:results} shows the performance differences between the three configurations for each load case. The Weibull averages over all load cases are shown in \cref{tab:weibull-performance}. The proposed controller reduces the rotor speed variations by $26\%$ with respect to regular BL control, such that the FOWT experiences only $4\%$ more rotor variations compared to operation in still water. 
These reductions take place in the wave frequency range, considered to be most difficult frequencies to attenuate by e.g. \cite{Lemmer_2016}.
As a side-effect of the novel controller, the power production is increased, the tower loads are reduced and the blade loads are reduced. These reductions require moderate additional pitch control action and result in slightly more shaft fatigue.
The FF controller is especially effective in severe environmental conditions because wave loads become more dominant over wind loads in more extreme load cases. This effect can be explained as follows: While the rotor thrust force decreases (because the blade pitch angle increases) for large wind speeds, the wave height increases. Therefore, the wave FF controller can reduce a larger percentage of rotor speed variations in severe environmental conditions.

\begin{table}[h]
    \centering
    \caption{Performance results, obtained from the high fidelity simulations for each controller and based on the Weibull average.}
    \label{tab:weibull-performance}
    \begin{tabular}{l|lll|l}
    \toprule
                          & BL    & BL+FF & BL (no waves) & $\frac{BL+FF}{BL}$ [\%] \\ \midrule
    mean $P$ [MW]         & 6090  & 6103  & 6106          & 0.21\%                  \\
    STD $\Omega$ [RPM]    & 0,32  & 0,23  & 0,22          & -26\%                   \\
    mean $\theta_c$ [RPM] & 0,032 & 0,13  & 0,0077         & 290\%                   \\
    DEL $M_{ty}$ [MNm]    & 64    & 55    & 12            & -13\%                   \\
    DEL $M_b$ [MNm]       & 3,9   & 3,3   & 2,9           & -15\%                   \\
    DEL $M_{LSS}$ [MNm]   & 0,55  & 0,57  & 0,42          & 3.4\%                   \\ \bottomrule
    \end{tabular}
\end{table}
   
\newpage
\section{Conclusion and outlook}
Based on high-fidelity simulations, it was shown that the novel feedforward (FF) control approach is able to significantly reduce wave-induced rotor speed variations, while indirectly reducing structural loads on the turbine and increase the energy capture. By complementing the regular feedback loop, the controller complexity is only increased by a linear transfer function and the regular stability properties are unaffected. Even though larger control actions are needed for nearly-full compensation, the framework allows a trade-off between stable power production and control input by using a simple proportional gain.

Whereas this work uses wave knowledge to regulate the rotor speed, the methodology here presented can be extended to attenuate two arbitrary system outputs, such as rotor speed together with platform pitch motion. Controllability of the control objective should be taken into account, as some control objectives may require large control actions. Future work will include the validation of the proposed technique via a wave basin test. The effects of second-order wave forces, which are found important for the dynamics of a floating offshore wind turbine, will be investigated as well. Furthermore, the effect of measurement errors and wind-wave misalignment will be reviewed.

\section{Acknowledgements}
This research has been partially funded by European Union through the Marie Sklodowska-Curie Action (Project EDOWE, grant 835901). 
\section*{References}
\bibliographystyle{iopart-num} 
\bibliography{main.bll}
\end{document}